\newif\iflak
\laktrue
\iflak
\documentclass{sig-alternate}
\else
\documentclass[a4paper,twoside]{article}
\fi
\usepackage[T1]{fontenc}
\usepackage[utf8]{inputenc}
\usepackage{epsfig}
\usepackage{subfigure}
\usepackage{calc}
\usepackage{amssymb}
\usepackage{amstext}
\usepackage{amsmath}
\iflak

\fi
\usepackage{amsthm}
\usepackage{multicol}
\usepackage{multirow}
\usepackage{pslatex}
\usepackage[backend=bibtex,hyperref=auto,
firstinits=true,style=authoryear,
urldate=iso8601,isbn=false,doi=false,
maxcitenames=1,maxbibnames=100]{biblatex}
\usepackage[hide]{ed}
\iflak\else
\usepackage{SCITEPRESS}
\fi
\usepackage[small]{caption}
\usepackage{float}
\usepackage{color}
\usepackage{hyperref}
\DeclareCaptionType{copyrightbox}

\iflak
\let\ednote\Ednote
\fi
\theoremstyle{definition}
\newtheorem{defn}{Definition}
\usepackage{enumitem}
\setlist[itemize]{topsep=0pt,itemsep=-1ex,partopsep=1ex,parsep=1ex}

\begin{document}

\title{OpenCourseWare Observatory -- Does the Quality of OpenCourseWare Live up to its Promise?}

\numberofauthors{3}
\author{
\alignauthor
Sahar Vahdati
\\
\affaddr{University of Bonn}\\
\affaddr{vahdati@uni-bonn.de}
\alignauthor
Christoph Lange
\\
\affaddr{University of Bonn}\\
\affaddr{math.semantic.web@gmail.com}
\alignauthor
Sören Auer
\\
\affaddr{University of Bonn}\\
\affaddr{auer@cs.uni-bonn.de}
}

\newcommand\ourkeywords{OpenCourseWare, Quality Metrics, Quality Assessment, Educational Content}
\iflak
\maketitle
\else
\keywords{\ourkeywords}
\fi

\newcommand\ourabstract{A vast amount of OpenCourseWare (OCW) is meanwhile being published online to make educational content accessible to larger audiences.
The awareness of such courses among users and the popularity of systems providing such courses are increasing.
However, from a subjective experience, OCW is frequently cursory, outdated or non-reusable.
In order to obtain a better understanding of the quality of OCW, we assess the quality in terms of \emph{fitness for use}.
Based on three OCW use case scenarios, we define a range of dimensions according to which the quality of courses can be measured.
From the definition of each dimension a comprehensive list of quality metrics is derived.
In order to obtain a representative overview of the quality of OCW, we performed a quality assessment on a set of 100 randomly selected courses obtained from 20 different OCW repositories.
Based on this assessment we identify crucial areas in which OCW needs to improve in order to deliver up to its promises.}
\iflak
\begin{abstract}
\ourabstract
\end{abstract}
\else
\abstract{\ourabstract}
\fi

\iflak
\category{K.3}{Computers and Education}{Computer Uses in Education}[Collaborative learning, Distance learning]
\keywords{\ourkeywords} 
\else
\onecolumn \maketitle \normalsize \vfill
\fi


\noindent\section{Introduction}

During the last decade the community of educators has been widely interested in improving the training model of education systems, towards high quality education \emph{in any place at any time}.
An important result of the collaborative work of educators and researchers in this direction is the OpenCourseWare (OCW) concept.
The idea arose from the success of open source software by expanding the concept of \emph{openness} to a larger context~\parencite{Vladoiu.2011}.

A vast amount of OCW is meanwhile being published online to make educational content more accessible.
The awareness of such courses among users and the popularity of systems providing such courses are increasing.
However, from a subjective experience, OCW is frequently cursory, outdated or non-reusable.
Many of the courses contain only a syllabus, are only available in one language or in formats, which are difficult to reuse.

More than 250 institutions worldwide are openly publishing courses today.
Some OCW initiatives by renowned universities are MIT OpenCourseWare, Stanford Engineering Everywhere,
Carnegie Mellon Open Learning Initiative, Harvard University Extension School, Open Learning Initiative,
Initiative, Open Yale Courses, Rice University’s OpenStax, OpenLearn, and Webcast.Berkeley.
Further OCW repositories have been made available by organizations such as OpenCourseWare Consortium, Open Education Resources Commons, and The Saylor Foundation’s Free Education Initiative.

The Open Education Consortium\footnote{\url{http://www.oeconsortium.org/}} as the central community of institutions and organizations working on open education lists 26,611 courses from 80 providers.
Many of the repositories mentioned above are members of the Open Education consortium.
MIT OpenCourseWare as one of the popular OCW repositories reports that they have made 2,150 courses available so far.
Since its launch, 137 million people have been visiting MIT OpenCourseWare annually.


The basic idea of OCW was to provide open access to educational material for educators, students, and individual learners around the world%
~\parencite{MIT2006}.
Instantly updated educational material should be freely available for everyone, or at least with lower costs, from anywhere at any time%
~\parencite{UNESCO.2002}.
Thus, OCW could form a big step towards achieving the right to education for everyone irrespective of race, gender, nationality, disability, religion or political preference, which is mandated by the Universal Declaration of Human Rights~\parencite{UnitedNations1948}.
OCW pioneers had the expectations that OCW would …

\begin{itemize}
\item help universities to attract prospective students from all around the world~\parencite{MITNews2001},
\item quickly disseminate of new educational content possible in a wide range of fields without waiting for academic publishers%
~\parencite{MITNews2001},
\item make quality material available in a variety of styles, languages and from a variety of viewpoints~\parencite{Caswell2008}.
\end{itemize}

The OCW pioneers promised to achieve these goals by constantly widening access to high quality digital educational materials.
To assess and improve the quality of OCW, a ``gold standard'' for reusable educational material first has to be established.
However, this task is not trivial, and one of the important challenges is a lack of representative and objective quality criteria.
It is proved, for example, by a large annual US national kindergarten to high school (K–12) survey~\footnote{\url{http://www.tomorrow.org/speakup/}}.
The results of 2011 showed that 41\% of principals find it difficult to evaluate the quality of digital content.
At the same time above 50\% of teachers responded that the most important factors in evaluating content were “being referred by a colleague”, “free”, and “created by educators”, none of which is necessarily a hallmark of quality \parencite{Porcello2013}.

We address this issue by establishing a set of \emph{quality metrics} for OCW.
Quality is defined as excellence, value, conformance to specifications, or meeting consumer expectations~\parencite{Kahn2002}.
More specifically, it is defined as \emph{fitness for use}~\parencite{Juran1974,knight2005}.
``Fitness for use'' means the extent to which the totality of features and characteristics of OCW leads to a successful fulfillment of its users' needs.
Our observatory will support or refute a preconceived subjective experience about the quality of OCW in terms of fitness for use by watching characteristics of courses.

In order to obtain a representative overview of the current state of OCW quality, we apply the quality metrics to observe the quality of a set of 100 randomly selected courses obtained from 20 different OCW repositories.
Based on this observation we identify crucial areas where OCW needs to improve in order to deliver up to its promises.
We introduce the methodology of this observatory in Section~\ref{sec:survey-methodology}.
Section~\ref{sec:metric-def} provides a concise definition of each dimension and of its quality metrics.
Section~\ref{sec:assessment-results} illustrates the results of our observatory.
Section~\ref{sec:background-work} discusses related work, and Section~\ref{sec:Con-Recom} concludes.

\noindent\section{Methodology}
\label{sec:survey-methodology}
We use a systematic observation as a structured, qualitative data collection and evaluation method.
Observation can be used to understand an ongoing process or situation~\parencite{Briefs.2008}, provide reliable, quantifiable data, or to collect direct information~\parencite{patton2005}.
Other sources on the Web also report that observation is to document detailed characteristics of objects
and apply a benchmark over a set of collected data.

Depending on the type of metric the observation is done as time or event sampling.
For example availability of course material from a server is studied in time intervals (see Section~\ref{sec:Availability}), whereas multilinguality is captured once (see Section~\ref{sec:Multilinguality}).
We first define three use case scenarios covering different OCW stakeholders.
Based on these scenarios, we introduce quality dimensions, including multilinguality, availability, discoverability.

For each dimension, we define quality metrics and justify their relevance.
For example, sustainability of a course is measured by the number of available revisions, their regularity over time, and their temporal distribution (see Section~\ref{sec:sustainability}).

To find courses, one can start with an initial set of widely known repositories (e.g. MIT OpenCourseWare), and further repositories from the list of members of the Open Education Consortium\footnote{\url{http://www.oeconsortium.org/members/}}.
Further courses can be retrieved using OCW-specific search engines:
1. There are authoritative listings of such search engines: one by the Higher Education Academy/JISC Open Educational Resources programme\footnote{\url{https://openeducationalresources.pbworks.com/w/page/27045418/Finding\%20OERs}} and one by the Open Knowledge Foundation\footnote{\url{http://booktype.okfn.org/open-education-handbook/}}.
2. From those search engines mentioned in both of these listings, we used those that were still available, and covered actual OCW repositories (rather than, e.g., Wikipedia), and covered multiple ones of them.
3. From these search engines, we obtained a list of courses.

Each of these ways allows for selecting a random sample of courses, which should be cleaned up to obtain a trustable and mature collection.
For example, courses with broken links or empty learning material should be disregarded.

At this point, the assessment process can be applied to each course by observing its characteristics w.r.t. the defined metrics.
The data resulting from this assessment should be recorded systematically to enable subsequent analysis.

\begin{table}[ht]
\centering
\scriptsize
\begin{tabular}{p{3cm}p{4.5cm}p{3cm}}
\hline
\textbf{Name} & \textbf{URL} \\
Connexions & \url{http://cnx.org} \\
Curriki & \url{http://www.curriki.org/} \\
JISC Digital Media & \url{http://www.jiscdigitalmedia.ac.uk} \\
Jorum & \url{http://www.jorum.ac.uk/} \\
Mellon OLI & \url{http://oli.cmu.edu/} \\
MERLOT & \url{http://www.merlot.org/} \\
MIT OpenCourseWare & \url{http://ocw.mit.edu/index.htm} \\
OCWFinder & \url{http://www.ocwfinder.org/} \\
OER Commons & \url{http://www.oercommons.org/} \\
OER Dynamic Search Engine & \url{http://edtechpost.wikispaces.com} \\
OpenCourseware Consortium & \url{http://www.ocwconsortium.org/} \\
OpenHPI & \url{https://open.hpi.de/} \\
Temoa & \url{http://www.temoa.info/} \\
The UNESCO OER Toolkit & \url{http://oerwiki.iiep.unesco.org} \\
TuDelft OpenCourseWare & \url{http://ocw.tudelft.nl/} \\
UCIRVINE & \url{http://ocw.uci.edu/} \\
University Learning & \url{http://www.google.com/coop} \\
Utah State OpenCourseWare & \url{http://ocw.usu.edu/} \\
webcast.berkeley & \url{http://webcast.berkeley.edu/} \\
Xpert & \url{http://xpert.nottingham.ac.uk/} \\
\end{tabular}
\caption{List of the OCW repositories (alphabetically sorted).}
\label{tab:oer-repo}
\end{table}

\noindent\section{Quality Metrics}
\label{sec:metric-def}

After analyzing OCW usage scenarios and doing a literature review, we identified a core set of 10 quality dimensions.
Dimensions are selected in a way that can be applied to assess the quality of OCW.
We group the identified dimensions according to the classification idea introduced by \citeauthor{Zaveri2012:LODQ} \parencite*{Zaveri2012:LODQ} as:
Accessibility dimensions [Availability, Discoverability], Intrinsic dimensions [Multilinguality level, Community involvement], Reuse dimensions%
 [Legal Reusability, Re-purposing format], Learnability dimensions [Learnability by examples and illustrations, Learnability by self-assessment], 
Temporal dimensions [Sustainability, Recency].

In the remainder of this section, we define each dimension in the context of OCW, and list metrics for measuring quality in this dimension.
We derive 37 quality metrics, including objective (O) and subjective (S) ones.
Our focus is on objective metrics, since they are better measurable and more reliable.
\autoref{tab:dimensions} provides a summary of dimensions, metrics and their definitions.
While almost all individual metrics have a numeric or Boolean value, we leave the interpretation, and possibly weighting, of these values to those who carry out a concrete assessment.
Additionally, a ``pros and cons'' section justifies the relevance of considering each dimension: it discusses the benefits of improving OCW quality w.r.t. this dimension but also points out possible challenges, obstacles and pitfalls in doing so.

\subsection{Legal Reusability}
\label{sec:Legal}
A large number of OCW users wants to build upon, enhance and (re)use the content of courses to reduce the effort of recreating material.
They need to be assured of the possibilities of legally reusing course content.
Therefore, each OCW should legally allow (re)use and adaptation of the content under an open license%
~\parencite{Friesen.2013}.
Several types of open licenses have been created, such as the Creative Commons licenses or the Open Publication License~\parencite{Atkins.2007}.
Each license specifies certain conditions, which can be combined with different sub-license attributes and types.
These certain conditions bring legal restrictions to protect the rights of each parties: original creator, sharing system and users.

According to the Creative Commons licenses\footnote{\url{http://creativecommons.org}}, we classify the conditions of reuse as follows:
\emph{Attribution (BY)} requires derivative works to give credit to the original creator and provide a link to the license.
\emph{Share-alike (SA)} requires derivative works to be distributed under a license identical to the original license.
\emph{Non-commercial (NC)} restricts (re)use of content to non-commercial purposes.
\emph{No Derivative Works (ND)} forbids derivative works from being published.

\begin{defn}
\emph{Legal reusability is the extent to which the terms and conditions specified by the creator grant the permission to legally (re)use content.}
\end{defn}

\noindent\textbf{Measuring:}
We measure legal reusability of a course by looking at its license.
When the course itself does not specify a license, we check whether the overall repository does so.
$M1.1$, a Boolean metric, is true if a license exists at all.
$M1.2$ indicates whether a human-readable description of a course's license is accessible from the web page of a course, be it that the page summarizes the license or links to the full definition of the license.
For each condition of reuse (BY, SA, NC, ND) we define three-valued metrics (false, true, unspecified), etc.
$M1.3_\mathit{BY}$, $M1.3_\mathit{SA}$, etc. specify the type of course license using these values.
We consider two separate metrics to measure the extent to which the license is machine-readable.
$M1.4$ measures whether a machine-readable indication of license exists,
and $M1.5$ indicates whether the description of the license itself is machine-readable.


\noindent\textbf{Pros:}
\begin{itemize}
\item License concisely summarizes the terms of reuse.
\item Permissive licenses grant more legal reuse possibilities.
\item Clear licensing conditions facilitate the content reuse (without cumbersome inquiries or negotiations).
\end{itemize}
\noindent\textbf{Cons:}
\begin{itemize}
\item Adding certain conditions to licenses can limit reuse.
\item Terms of a license might be difficult to understand or require interpretation and adaptation in certain legislations.
\item In practice, it is difficult to track whether material is being reused in accordance with its license.
\end{itemize}

\begin{table*}[tb!]
\scriptsize
\centering
\begin{tabular}{p{3.5cm}p{10.5cm}r}
\hline
\emph{Dimension} & \emph{Metric} & \emph{Type} \\ \hline
\multirow{5}{3cm}{M1. Legal reusability} & {\small M1.1 Existence of license for a course} & O \\
& {\small M1.2 Existence of human-readable description of license} & O \\
& {\small M1.3 Type of legal (re)usability} & O \\
& {\small M1.4 Existence of machine-readable of license} & O \\
& {\small M1.5 Existence of machine-readable description} & O \\	
\hline
\multirow{4}{*}{M2. Multilinguality level} & {\small M2.1 Identification of the original language} & O \\
& {\small M2.2 Existence in other languages} & O \\	
& {\small M2.3 Number of further language in which a course is available} & O \\
& {\small M2.4 The state of translation: automatic, synchronized, expert-revised, localized} & O \\
\hline
\multirow{3}{*}{M3. Format re-purposeability} & {\small M3.1 Format of the course material} & O \\
& {\small M3.2 Possibility for reuse} & O \\	
& {\small M3.3 Type of function for reusability} & O \\
\hline
\multirow{2}{*}{M4. Recency} & {\small M4.1 Average recency of individual modules and content units} & O \\
& {\small M4.2 Recency of the overall course} & O \\
\hline
\multirow{3}{*}{M5. Sustainability} & {\small M5.1 Number of available revisions} & O \\
& {\small M5.2 Regularity of a course versions over the lifetime of the course} & O \\
& {\small M5.3 Average recency of revisions} & O \\
\hline
\multirow{5}{*}{M6. Availability} & {\small M6.1 Server’s availability} & O \\
& {\small M6.2 Presence of the material} & O \\
& {\small M6.3 Availability of the content for download} & O \\
& {\small M6.4 Portability of a course on different devices with different operating systems} & O \\
& {\small M6.5 Availability of the format and structure of the content on different devices} & O \\
\hline
\multirow{5}{3cm}{M7. Learning by self-assessment} & {\small M7.1 Existence of self-assessment material} & O \\
& {\small M7.2 Mean number of self-assessment objects in a course} & O \\
& {\small M7.3 Coverage of self-assessment material over the course} & O \\
& {\small M7.Sol.1 Existence of solutions for self-assessment material} & O \\
& {\small M7.Sol.2 Mean number of self-assessment solution objects in a course} & O \\
\hline
\multirow{3}{3cm}{M8. Learning by examples and illustrations} & {\small M8.1 Number of examples over the total number of course units} & O \\
& {\small M8.2 Number of illustrations over the total number of course units} & O \\
& {\small M8.3 Attractiveness level of a course} & S/O \\
\hline
\multirow{6}{3cm}{M9. Community involvement} & {\small M9.1 Type of course creation: single author or collaboration work} & O \\
& {\small M9.2 Number of contributors for the courses} & O \\
& {\small M9.3 Number of learners or educators} & O \\
& {\small M9.4 Number of comments written by users} & O \\
& {\small M9.5 Number of times that the course material is being downloaded by users} & O \\
\hline
M10. Discoverability & {\small M10.1 Average rank of a targeted course retrieved in the search result} & O \\ \hline
\end{tabular}
\caption{Overview of OCW quality dimensions and their metrics.}
\label{tab:dimensions}
\end{table*}

\subsection{Multilinguality Level}
\label{sec:Multilinguality}
The mission of OCW is to provide education for anyone at any time in any place.
However, the access to produced content is often limited by language barriers.
83 different languages are spoken by more than 10 million native speakers each.
Out of an estimated 2 billion Internet users, some 27\% percent speak English.
As speakers of other languages get online, the share of English speakers is decreasing.
Thus, the need of providing OCW translation in languages other than English is apparent.

In the context of education, the author's priority in turning a course into a multilingual one is to provide high quality while keeping the effort for translation low.
The following technologies help with this: (1) machine translation, (2) synchronization,
(3) internationalization and localization.
Machine translation can support manual translation, but the quality of output is still far below human translation%
~\parencite{Green.2013}.
An initial machine translation can help to reduce the effort by about a half but humans have to review and revise the output in order to reach a good translation quality.
After a course has been translated, it is important to keep improvements to the original version synchronized with improvements to the translated version.
Localization is the adaptation of the translated versions to cultural differences.
Examples are units of measurements (e.g., inch vs. centimeter), religious and regional differences.

\begin{defn}
\emph {Multilinguality means availability of material in multiple languages.}
\end{defn}

\noindent\textbf{Measuring:}
We consider a course multilingual whose content is available in more than one language.
Every course is associated with at least one language.
The chronological first language in which the course was designed is recorded as the original language.
$M2.1$ is defined as the original language.
$M2.2$ is a Boolean metric telling whether a course is available in different languages.
$M2.3$ records the number of further languages.
$M2.4$ specifies the state of the translations, which can be:
\begin{itemize}
\item \emph{automatic-translation} when a course is machine-translated and not reviewed by human experts.
\item \emph{synchronized} when the verbatim translation of the original language is edited to be synchronized with the new language.
\item \emph{expert-revised} when the translation was reviewed by a domain expert but not yet by native speaker.
\item \emph{localized} when a translated version of a course is checked by native speakers and localized.
\end{itemize}

\noindent\textbf{Pros:}
\begin{itemize}
\item Multilinguality reaches a wider audience and ensures wider usage of the material.
\item Multilinguality reduces the effort of material creating.
\item Localization addresses cultural differences.
\end{itemize}
\noindent\textbf{Cons:}
\begin{itemize}
\item Translation can be time-consuming and expensive.
\item Translation must be performed or carefully checked by domain experts.
\item Scientific or technical content needs to be adapted to the respective cultural context.
\end{itemize}

\subsection{Format Re-purposeability}
\label{sec:Format}
The content of a course can be reused by different groups of users for several purposes.
While an educator might want to reuse the content of a course in his/her lecture, a student might reuse the same content for self learning.
The format and its granularity can also influence the accessibility of the content.
For example, audio and video formats can only be edited in a very limited way, in contrast to ePUB or HTML.

Courses have been made available in different formats, such as interactive documents, audio, and video.
Interactive documents are web-based objects with lightweight interactive elements, e.g., for navigation or question answering; they can be implemented using HTML5/JavaScript or Flash.
Text can come in different formats such as HTML, ePUB, XML, PDF and plain text.
Representation of mathematical formulas is possible in \LaTeX{} or ePUB editors.
But even then it is problematic to copy and paste them for later reuse.
Copy-paste from PDF or even certain text files can cause errors while copying special characters.
Simulations are a different format that are usually available for certain technical software.
Depending on the format re-purposing can be impossible or subject to restrictions (such as loss of presentation quality or of formatting during copy-paste).

The choice of formats not only influences re-purposing but also viewing.
Some users might not be able to read the content because of certain technical requirements such as the need to install a certain software (e.g., Flash).
Therefore, accessibility and re-usability of the format are key requirements.

\begin{defn}
\emph {The term ``re-purposing'' is used when the usage purpose of the content changes depending on the target audience.
A re-purposeable format gives direct access to the course content with no or minimal loss of information.}
\end{defn}

\noindent\textbf{Measuring:}
$M3.1$ represents the format of the course material.
$M3.2$ is a Boolean value indicating whether the content is reusable (for example Video is not, PowerPoint and HTML is).
$M3.3$ indicates how course content can be reused.
Values of the metric are the possible functions for reuse e.g., copy/paste function or by direct editing.

\noindent\textbf{Pros:}
\begin{itemize}
\item Re-purposable format enables technical openness.
\end{itemize}
\noindent\textbf{Cons:}
\begin{itemize}
\item Sufficiently re-purposable formats are rarely available.
\item Format reusability restrictions can conflict licenses.
\end{itemize}

\subsection{Recency}
\label{sec:Recency}
Learners are interested in courses reflecting the state of the art.
Therefore it is important to study temporal aspects of OCW.
A course that was good in the past may not be a good course now or in the future.
If we consider OCW to be the digital reflections of courses taught in reality at a certain time, their age and content freshness becomes a relevant quality indicator.

Frequent updates can keep the users of a course satisfied with its freshness.
This can influence the popularity of a course as well as ranking in time sensitive retrieval via search engines~\parencite{Dong.2010}.
Apart from constant facts and proved theories, scientific content carried by OCW could require updates over time.
Therefore, recency of OCW depends on the awareness of its instructors of the changes of the concepts over time.
Not only modifications of the content should be considered, but the means of representation can also be improved over time, thus influencing the attractiveness of a course.

\begin{defn}
\emph{Recency is the extent to which the content and the material of a course is updated.}
\end{defn}

\noindent\textbf{Measuring:}
Unit is the granularity in which each leaf of course content has been made available, e.g., page, slide, or interactive document.
OCW recency can be measured on two levels: the average recency of individual content units of the course, and the recency of the overall course.
$M4.1$ depicts the average recency of a course w.r.t. updates over individual parts of a course.
$M4.1_1$ measure the recency of course modules and $M4.1_2$ considers recency of content units. 
$M4.2$ shows the recency of the overall course.
Recency is defined as the difference between the date of measurement $t_{\mathit{obs}}$ and the date when a course was last updated $t_{\mathit{lastUpd}}$ (cf. \parencite{BaezaYates.2004}).
It is not a very precise measure, since sometimes only minor updates have happened.
In our evaluation we measure with a granularity in years because in our 100 sample courses (cf. section~\ref{sec:assessment-results}) we observed that most courses are taught and updated once a year.

\noindent\textbf{Pros:}
\begin{itemize}
\item Recency is a requirement for covering the state-of-the-art.
\item Recency is an indicator for sustainability and popularity.
\end{itemize}
\noindent\textbf{Cons}:
\begin{itemize}
\item Recency is difficult to measure on a content unit basis -- requires some form of revision control.
\item Typically recency of OCW does not have disadvantages for users.
Except if old versions of the course material were not archived (cf. section~\ref{sec:sustainability}), then by updating a course and, e.g., deleting a section, one would lose old information.
\end{itemize}

\subsection{Sustainability}
\label{sec:sustainability}
An important challenge for projects aiming at free education is their economic sustainability over time.
In~\parencite{Dholakia.2006}, sustainability is introduced as the long-term viability and stability of an open education program.
\Citeauthor{Downes.2007} categorize the sustainability of Open Educational Resources (OERs) from three point of views: funding models, technical models and content models.
He considers the sustainability of open educational resources to meet provider's objectives such as scale, quality, production cost, margins and return on investment.

These definitions consider the sustainability of OER projects both from a commercial and a technical point of view.
In this article, we focus on the sustainability of OCW from a content perspective.
In most cases, a course is not taught once but rather multiple times over several semesters or years.
Although instructors aim at a consistent style and content in their courses, small refreshments are always necessary.
These changes can be either infrastructural editions in the whole content or slight updates in sentences, paragraphs.
By each edition, a new variant of a course is created, which could be managed using a version control system.

Sustainability of OCW projects and systems depends on many external factors, such as funding, objectives of stakeholders, awareness of users, advertising, etc.
We do not include these aspects of sustainability in this survey because they do not apply to courses.

\begin{defn}
\emph {Sustainability of OCW shows their quality from the aspect of being stable over time.
The quality of being stable is defined by the amount of previous versions and their regularity over time.}
\end{defn}

\noindent\textbf{Measuring:}
A long and continuous revision history shows that the content of a course has been well maintained in the past.
This indicates that it may also continue to be well maintained in future.
Some OCW repositories offer a revision history of the courses.
Using this information, we measure sustainability of a course by $M5.1$ the number of available revisions.%
\footnote{We refer to the change that transformed version $n-1$ to version $n$ as a \emph{revision}. A revision occurs at a precisely defined point in time.}
While a high \emph{number of revisions} indicates that the authors of a course have devoted a lot of attention to it, it is not reliable to measure the sustainability of a course only by counting the number of revisions.
Therefore, two attributes are considered while measuring sustainability of learning objects:
$M5.2$ indicating the regularity of a course's versions over the lifetime of the course and
$M5.3$ measuring the average recency of all revisions.

We define revisions of the courses to be regular if any two successive revisions have the same time difference.
This notion of regularity is valid only for courses with more than two revisions.

Apart from regularity, the recency of versions is also important (see Section~\ref{sec:Recency}).
The \emph{recency} of versions is calculated as the variance of their distribution over the lifetime of a course.
A high variance indicates that the versions of a course tend to be updated frequently, while a low variance indicates that the versions have been updated a long time ago.

\noindent\textbf{Pros:}
\begin{itemize}
\item A course with a continuous history of old versions enables users to understand how concepts, their definitions, and their explanation have evolved.
\item Giving users access to previous versions gives them the possibility to study the evolution of the content from the beginning until the most recent update.
\end{itemize}
\noindent\textbf{Cons:}
\begin{itemize}
\item Limiting access to a single version, i.e.\ the most recent one, prevents users from understanding the evolution of the content of the courses.
\item It is a difficult task for users to realize the exact changes of the content in each unit without version control facilities
\item Assisting users with version control features depends on the technical capabilities of an OCW repository engine.
\item Regularity and recency of versions directly depends on the contribution of authors.
\end{itemize}

\subsection{Availability}
\label{sec:Availability}
A generalized definition for availability is given by~\parencite{Katukoori.1995} as the ratio of times that a resource is capable of being used over the aggregation of uptimes and downtimes.
In the context of the Web, Bizer defines availability as the extent to which information is ``available in an easy and quick way''%
~\parencite{Bizer2008:PhDThesis:biblatex}.
There are various ways of making OCW available, i.e. ready for use to its users.
These include: making the course available as a website, 
making a specific part (i.e. unit) of the course available (and shareable) via a URL,
making the course available in a content repository (e.g. a video/course archive),
offering a whole course for download in various formats (e.g. PDF, presentation),
offering individual learning objects for download.

These different ways are not mutually exclusive.
For example, a course can be made available as a website as well as an archive for download through some content repository.
The possibility to download a course makes it available for students who do not have permanent access to the Internet.

Different formats in which course material is offered can be distinguished by their portability, size and accessibility.
A format is portable if (free) software for viewing it without losses is available for all major operating systems.
Different formats may result in different download sizes, which matters when a user's internet bandwidth is low.

Finally, different formats have different degrees of accessibility not just for people with disabilities, but also for users of devices with small screens low screen resolutions (smartphones).
We define accessibility as the extent to which the format of a course makes it available to a broad range of users.

In most formats, there are ways to increase the level of accessibility.
For example, closed captions in video lectures display transitive text information that can be activated by the viewer.
Another important aspect is whether the users are able to download the courses instantly, or whether they have to create an account and log in.
Adding such a registration barrier can conflict with the meaning of `open'.

\begin{defn}
\emph {Availability of OCW is the extent to which its learning objects are available in an easy and quick way.}
\end{defn}

\noindent\textbf{Measuring:}
We define five measures of availability concerning different aspects of OCW usage.
$M6.1$ measures the server's availability, %
$M6.2$ indicates the presence of the material, %
$M6.3$ measures factors concerning the availability of the content for download and %
$M6.4$ characterizes the portability of a course on different devices with different operating systems and
$M6.5$ indicates attributes related to the format and structure of the content.

$M6.1$ is calculated as the ratio between the number of times the server was available over the number of checking times.

$M6.2$ indicates whether all parts of an OCW are available (not necessarily for \emph{download}; %
see $M6.3$ below).
It is a Boolean measure, which is \emph{false} in case of incompletely available course material or its absence, and \emph{true} otherwise.
During our study we faced cases when only the metadata of a course was available, whereas in other cases some course parts were missing; in both cases %
$M6.2$ would take the value \emph{false}.
For video lectures, $M6.2_1$ indicates whether the course content is facilitated by closed caption process.

$M6.3$ measures factors concerning the availability of the content for download.
When a course is available for download, users can either download the whole course material at once or every part or chapter has to be downloaded as an individual file.
We consider both possibilities to be important for availability and therefore define two Boolean sub-metrics, %
where $M6.3_1$ indicates the downloadability of the whole course at once, %
and $M6.3_2$ indicates the downloadability of all of its parts.

$M6.4$ comprises three independent Boolean sub-metrics to measure the portability of a course.
$M6.4_1$ measures whether the material is available in a format for which viewer applications are available for all major operating systems\footnote{As major operating systems we consider Windows, Mac OS and Linux (as their combined market share on PCs is 93.7), as well as iOS and Android (as their combined market share on mobile devices is 94.4).}
(Example: videos that require Flash, which is not available on Android).
$M6.4_2$ indicates whether the material is available in a format that can be viewed \emph{without losses} on all major operating systems.
(Example: There is software for all major operating systems that can view PowerPoint, but only \emph{Microsoft} PowerPoint, which, e.g., is not available for Linux, can view PowerPoint documents without losses).
$M6.4_3$ indicates whether the material is available in a format for which \emph{free} viewer applications are available for all major operating systems.
(Example: Microsoft PowerPoint is available for Windows and Mac OS, but it is not free.)

$M6.5$ is a Boolean metric to measure the availability of the content structure.
$M6.5_1$ depicts whether the content is easily available in smaller granularity, for example in that the all-in-one archive contains a table of contents, or by having multiple archived files for download, e.g.\ one per chapter.

\noindent\textbf{Pros:}
\begin{itemize}
\item Availability is a necessary condition for openness.
\item Having a course available in smaller granularities gives the advantage of easy access to the desired content.
\item Availability of learning objects in downloadable formats ensures that users will always be able to access material.
\end{itemize}
\noindent\textbf{Cons:}
\begin{itemize}
\item Availability is influenced by several independent preconditions; for example: a course with a smartphone-friendly online document format is effectively not available while the web server is down.
\item For a student it is a laborious task to download a complete course if the material is only available as multiple separate archives.
\end{itemize}

\subsection{Learnability by Self-assessment}
\label{sec:SA}
Learning can only be effective if the learner is aware of the progress made so far, and of the amount of remaining knowledge that needs to be covered~\parencite{Boud.1995}.
Self-assessment is an old technique which is mostly used in order to foster reflection on one's own learning processes and results~\parencite{Dunning.2004}.
Using OCW, there is no human instructor to tell the learner what to do or how to proceed.
Therefore, self-assessment plays an important role in helping the learner to reach an optimal level of knowledge.
Learning material for self-assessment can be categorized in three different classes: 
(1) exercise sheets for training, 
(2) exam sheets that help with exam preparation, and 
(3) quizzes or multiple choice questions, helping to measure learning progress.

\begin{defn}
\emph{Self-assessment material enable learners to assess and improve their knowledge.
As a quality attribute, we define the extent to which a course supports its audiences (users) in understanding the contained content by offering self-assessment material.}
\end{defn}

\noindent\textbf{Measuring:}
In this work, different parts or chapters of a course where instructors partition the whole course is called ``module''.
Number of modules for a course is denoted as $N_m$.
We count self-assessment objects as the number of individual objects e.g., one exam sheet with 10 exercises counts as ``10'' rather than ``1''.
First, with $M7.1$ as a Boolean metric, we check whether any kind of self-assessment material exists for a course at all.
Then, for a course module $i=1,\dots,N_m$ we denote the number of self-assessment objects for this module as $\mathit{sa}_i$; thus, the overall number of self-assessment objects in a course is $N_\mathit{sa}:=\sum_{i=1}^N \mathit{sa}_i$.

$M7.2$, abbreviated as $\mu_\mathit{sa}$ in this section, is the mean number of self-assessment objects in a course, i.e.\ the number of self-assessment objects divided by the number of course modules ($M7.2 =\mu_\mathit{sa}:=\frac{N_\mathit{sa}}{N_m}$).

Furthermore we are interested in the statistical distribution of self-assessment objects over course modules.
We consider modules rather than units for self-assessment because of high possibility of their distribution over modules.
Note that the relation of self-assessment objects to course modules may not always be easy to determine: in some courses, self-assessment objects are attached to modules; however, if a course only has one overall self-assessment block at the end, which applies to all modules, determining this relation will require linguistic or semantic analysis of the self-assessment objects and the content of the course modules.
We leave the decision of whether to determine the distribution of self-assessment objects over course modules, or to leave the respective metrics undefined, to the ``user'' of these metrics definitions (vs., e.g., the ``end user'' = the student using OCW material for learning).

$M7.3$ is defined as the ratio of course modules with at least one self-assessment object to the overall number of modules.
This definition is inspired by code coverage in software testing.

For any self-assessment object a solution or answer may or may not be included.
Therefore, for each metric $M7.x$ defined above, we define another metric $M7.{\mathit{Sol}}.x$, which only takes into account objects having solutions.
For example, $M7.{\mathit{Sol}}.1$ is true if there exists self-assessment material with solutions.
Coverage is not measured for solutions as it can be determined by self-assessment objects.

\noindent\textbf{Pros:}
\begin{itemize}
\item Self-assessment material are useful for checking one's understanding of the key messages and points about a subject.
\item Self-assessment is necessary for effective learning.
\item Having exam sheets available for users or recommending extra reading material can reduce the need for further searches.
\item Having self-assessment material attached to a course gives it a wider range of users.
\end{itemize}
\noindent\textbf{Cons:}
\begin{itemize}
\item It is very difficult to find a pool of self-assessment material.
\item The difficulty of self-assessment exercises should match the difficulty of the corresponding learning object; otherwise the results can be unreliable.
\item Self-assessments can be effective when the learners understand the solutions.
\item A certain level of knowledge is required to train the self-assessment exercises.
\end{itemize}

\subsection{Learnability by Examples and Illustrations}
\label{sec:illustration}
Instructors commonly place different adjunct items in the course content to facilitate learners understanding from the underlying subjects%
~\parencite{Hayes.1983}.
They often choose illustrative items rather than pure text descriptions that can be easily seen, remembered and imagined.
Illustrations refer to any kind of graphical representations such as graphs, pictures, videos.
Reviewing studies about ``reading-to-learn'' concludes that at least well-selected and well-instructed items can reliably improve learning%
~\parencite{Carney.2002}.
In some cases using pictures might not increase learning rather makes the course material attractive.
Apart from pictorial items, examples as single and specific instances are frequently used to reveal complex concepts.

\begin{defn}
\emph {The degree to which the content of an OCW is well-illustrated by examples and illustrations shows its level of learnability.}
\end{defn}

\noindent\textbf{Measuring:}
We consider the use of examples as well as illustrations within a course content that are used to convey complex ideas and messages.
$M8.1$ is calculated as the ratio between the number of examples over the total number of course units.
Examples are counted as the instances titled by the term ``example''.
Similarly, $M8.2$ specifies the ratio between the number of illustrations and the total number of course units.
Any kind of instances than pure text such as pictures, charts, graphs, diagrams, and tables are counted as illustrations.
Summation of these two measures specifies the level of course learnability by examples and illustrations.
As mentioned before, the number of illustrations effects the attractiveness level of a course.
$M8.3$ measures the level of course attractiveness based on two sub-metrics.
$M8.3_1$ is the ratio between number of illustrations over the total number of course units.
$M8.3_2$ is a subjective factor to determine the level of attractiveness.  

\noindent\textbf{Pros:}
\begin{itemize}
\item Concepts with additional illustrations and examples improve performance of learning.
\item Illustrations increase motivation and creativity of learners.
\item Examples and illustrations increase attractiveness level of a course.
\item They reinforce the text’s coherence and give supports for readers in text-processing. 
\end{itemize}
\noindent\textbf{Cons:}
\begin{itemize}
\item Illustrations are sometimes exact representation of what is described in the text or the other way around.
\item Attractiveness is a subjective criterion.
\end{itemize}

\subsection{Community Involvement}
\label{sec:Involvement}
Many content projects are based on collaborative work of individuals who put their time, effort and skills to creating a product that is available to all~\parencite{Oreg.2008}.
Usually a course has one primary author as the instructor of the course who does the major work and few others of doing minor contribution.
Two kind of collaboration is possible creating OCW: collaboration in the background and online collaboration.
This means whether the revisions of the content are created by community in the background and uploaded by one of the authors.
Second, the system can make it possible to have a collaborative environment and work on one unit with many users.
Two groups of people can collaborate to create learning objects: members of a group, volunteers.
Creating learning objects is achievable even without collaborative work.


\begin{defn}
\emph {Collaboration on creating OCW is an interpersonal process through which members of same or different disciplines contribute to a common learning object.}
\end{defn}

\noindent\textbf{Measuring:}
$M9.1$ measures whether the OCW is created in a collaboration process or it has one author.
$M9.2$ measures the number of contributors for the courses created or edited by several people.
$M9.3$ shows the number of learners or educators who used the course and $M9.4$ depicts the number of comments written by users.
$M9.5$ describes the number of times that the course material is being downloaded by users.
In many cases we end up with situations that less information were available for these.

\noindent\textbf{Pros:}
\begin{itemize}
\item Collaboration can increase the freshness level of a course.
\item An OCW created from a collaboration work can be completed from several aspects that other courses with only one author.
\item A collaborative work can save a lot of time for authors and let them be more creative in the content.
\end{itemize}
\noindent\textbf{Cons:}
\begin{itemize}
\item Without a intellectual control the main objectives of the course may be lost.
\item Editions are not always productive.
\item Revision control is an essential need for a synchronized collaboration work.
\end{itemize}

\subsection{Discoverability}
\label{sec:Discoverability}
Although a large amount of courses is being published as OCW by universities and organizations, finding relevant educational content on the Web is still an issue~\parencite{Dichev.2011}.
Selection of certain results by users depends on the results shown by the search engine.
Regardless of being high or low quality OCW, their rank can be influenced by several factors which parts of them are far from the scope of this paper.
In this paper, we consider how easy relevant OCW are discoverable for users in the retrieved results of their browsing.

Discoverability refers to user's ability to find key information, applications or services at the time when they have a need for it%
~\parencite{Vladoiu.2013}.
It is the task of repository administrators to optimize their content’s discoverability.
They need to add certain factors behind the scenes of the website to improve content discoverability of their website by search engines.
The attempt of discovering courses among the search results has been done assuming that users are unaware of existence of the exact course.
The title of each course is normalized  and used as the search keyword.
Due to diversity of retrieved results only by titles, the search keyword is enriched by adding terms ``course'' and ``open''.

\begin{defn}
\emph {Discoverability of OCW is the extent to which it is promptly possible for users to find relevant courses ones to their search at the time of need.}
\end{defn}

\noindent\textbf{Measuring:}
Two factors can directly influence the search results while searching the web: the search engine and the search keywords.
Users searching for courses have at least basic knowledge about them including a list of search keywords.
Combination of each keywords differently feed the search engine discovering courses.
Different metadata of courses such as title, topic, author, etc are used as search keywords.
Furthermore, adding phrases like ``course'' or ``open'' can influence the result retrieved from search engines.
In our assessment, we use Google (with 1,100,000,000 estimated unique monthly visitors) and being the top as the test search engine measuring discoverability of courses.\footnote{\url{http://www.marketingcharts.com/}}
During last four years, over 87 percent of worldwide internet users searched the web with Google search engine.\footnote{\url{http://www.statista.com/statistics/216573/worldwide-market-share-of-search-engines/}}
First we used Google's advanced search to narrow down the number of retrieved results to 100.

$M10.1$ measures the average rank of a targeted course retrieved in the search result.
By taking samples from the collection we can see how discoverable are the courses among first 100 results shown by the Google search engine.
The results are influenced by using ``AND'', ``OR'' between different phrases.

\noindent\textbf{Pros:}
\begin{itemize}
\item Discoverability increases users' awareness of the existence of relevant repositories and courses.
\item When a good course is easily discoverable, it saves a lot of time for learners.
\end{itemize}
\noindent\textbf{Cons:}
\begin{itemize}
\item Making OCW easily discoverable on the Web is a challenging task for repository administrators.
\item A good course which is not easily discoverable, indirectly restricts users freedom of access and usage of its content.
\item Key factors influencing discoverability of OCW are out of author's authority.
\end{itemize}

\noindent\section{Assessment and Results}
\label{sec:assessment-results}
The main objective of this paper is to assess the quality of OCW.
We assess the quality of individual courses, not of repositories.
The study is based on 100 courses randomly selected from the 20 repositories shown in \autoref{tab:oer-repo}.
The repositories were chosen by including a mix of renowned OCW initiatives such as MIT OpenCourseWare or OER Commons and less prominent initiatives.
Now we discuss the outcome of the quality assessment of these courses according to the metrics defined in the previous section.
The results of the assessment for each dimension is summarized first and possible recommendations are given thereafter.\footnote{The full raw data and some visualizations are available at \url{http://bit.ly/1vhaWJT}.}

\noindent\textbf{Legal Reusability:}
All the courses in our sample collection are licensed.
Overall 28 out of the 100 courses have an open license.
Creative Commons (CC) is a ``family'' of licenses, out of which CC-BY-NC-SA (50 out of 100) is the most popular one.
However, because of the restriction to non-commercial reuse, CC-BY-NC-SA not an open license according to the Open Definition.
Overall 57 courses are licensed as non-commercial where as 3 courses out of 100 granted as non-derivative.
For most of the courses with a CC license, a human readable version of the license has been made available. 
License information in a machine readable format is provided for a small number of courses(20).

Almost all repositories use standard licenses with the exception of the OpenHPI repository.
In OpenHPI courses are licensed by a set of rules that the repository maintainers call ``Code of Honor'', unless particular course authors override these with their own license.

\noindent\textbf{Multilinguality Level:}
Of the four metrics defined to measure the level of multilinguality, it was not possible to measure \ednote{refer to the metric}the state of translation, since the required information was not provided by the repository.
While English is the original language of the majority of the courses
, two of them have been translated to other languages.
Out of 12 courses originally offered in other languages than English, four have been translated to English.
None of the repositories in our assessment offers real-time machine translation functionality.
As \autoref{tab:multilingual} indicates, English dominates the OCW realm by several orders of magnitude, while courses in other languages would be in high demand considering the number of internet uses.

\begin{table}[ht]
\begin{tabular}{p{.15\linewidth}p{.32\linewidth}p{.42\linewidth}}
\hline
\textbf{Language} & \textbf{Number of Courses} & \textbf{Internet Users} \\ \hline
English & 88 & 536 million \\
Chinese & 1 & 444 million \\
Spanish & 4 & 153 million \\
Japanese & 1 & 99 million \\
Portuguese & 1 & 82 million \\
German & 1 & 75 million \\
Others & 4 & <1 million \\
\end{tabular}
\caption{Number of courses and Internet users by language.}
\label{tab:multilingual}
\end{table}
 
\begin{figure}[h]
\includegraphics[trim=0.20cm 1.80cm 0.20cm 2cm,clip,width=\linewidth]{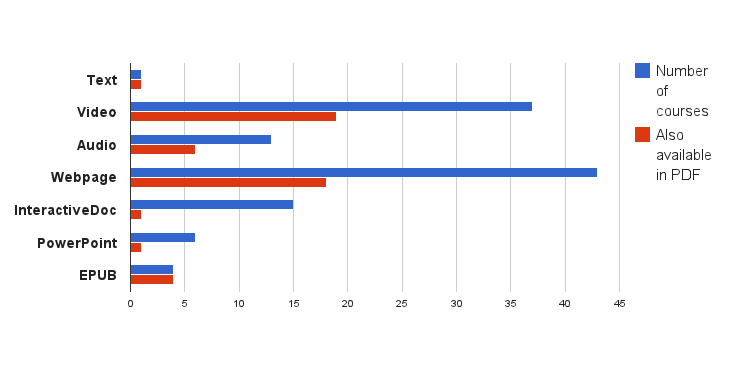}
\caption{Course formats}\label{fig:re-purpose}
\end{figure}

\noindent\textbf{Format re-purposeability:}
Overall 68 courses are offered in re-purposeable formats.
However, the problematic part is the way of their re-usability.
A large number of courses (52) are available in PDF and it is only possible to re-use content using copy-paste functions in a cumbersome way.
\autoref{fig:re-purpose} shows the number of courses w.r.t. the formats in which they are available.
The red bar shows the number of courses available in the corresponding format and additionally in PDF.

\noindent\textbf{Recency:}
A description of updates per module was only available for two courses.
In most of the cases the OCW repository software doesn't support it, and for most of supported ones the course authors didn't provide sufficient metadata.
Therefore, we consider the recency of the overall course by calculating the difference between the observation date (2014) and last update of the course.
Out of 100 courses, 10 have been updated in 2014
.
Overall, only 32 out of 100 courses have been updated in 2014 or in the two previous years.
More than half of the courses were last updated three years ago or earlier.
11 courses did not provide specific information about their last update.
These were mainly interactive documents with course contents. 

\begin{figure}[h]
\includegraphics[width=\linewidth]{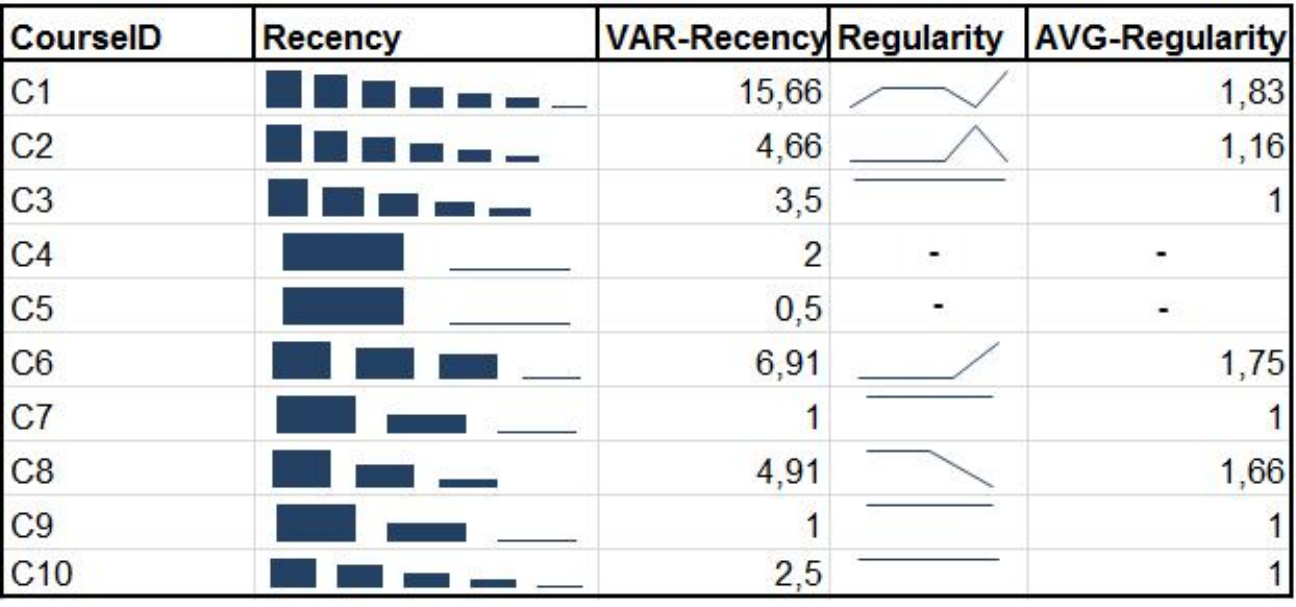}
\caption{Sustainability of courses}
\label{fig:sustainability}
\end{figure}

\noindent\textbf{Sustainability:}
Information about the revision history is only available for 14 courses in our sample set.
Four of these, however, were revised only within a single year.
\autoref{fig:sustainability} shows the sustainability of the remaining 10 courses w.r.t. average recency and regularity of revisions over the lifetime of each course.
The regularity of courses is depicted as line charts where the X axis represents the revisions and the Y axis shows the recency of the revisions.
Course number 1 from Webcast.Berkeley with seven revisions is the most sustainable course in our data set.
It has the highest of number of revisions, the highest recency variance and the highest average regularity.

Course 3 and course 10 with five revisions and average regularity of 1 are also sustainable over time.
Course 10 with less average recency of revisions is more sustainable than course 3.
We can not determine the regularity of courses number 4 and 5 since they do not satisfy the prerequisite of having more than two revisions.
Overall only four out of the ten course for which a revision history was available were sustainable according to our definition.

\noindent\textbf{Availability:}
Availability of servers has been checked in three time intervals.
In the second round of checking, 5 courses could not be accessed because of server problems.
This number decreased to 2 in the third round.
Some repositories restrict access to the course material by requiring an account.
For example, the Curriki repository limits the access to a maximum of three courses without an account.
\begin{figure}[h]
\includegraphics[trim=0.10cm 2.5cm 0.20cm 2.5cm,clip,width=\linewidth]{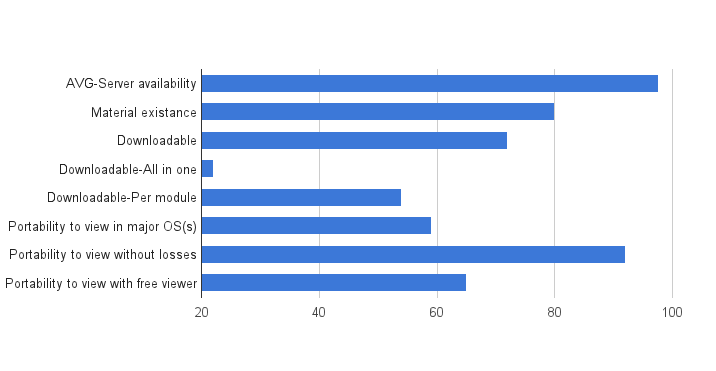}
\caption{Availability of courses}
\label{fig:availability}
\end{figure}
Eight courses are available in PowerPoint format.
Although there are open viewers like LibreOffice for operating systems where PowerPoint is not available (such as Linux), the formatting of these courses looks broken in parts.
Overall 18 courses out of 37 in video format have closed captions. 
Half of the courses provide the content in a structured format to download. 
10 courses out of 22 which are downloadable all-in-one contain a table of contents.
Only four courses are offered for download all-in-one and per chapter.
The content of 40 courses which are downloadable per chapter is archived in multiple files.
 
\noindent\textbf{Learning by Self-assessment:}
Self-assessment material is available separately for 40 courses.
15 courses include self-assessments directly inside the content. 
Out of 55 courses with self-assessment 25 of them have solutions. 

\begin{figure}[h]
\includegraphics[trim=0cm 0cm 0cm 0.10cm,clip,width=\linewidth]{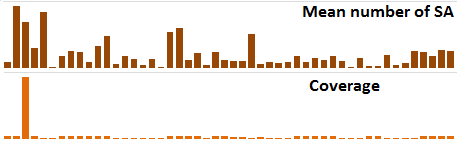}
\caption{Self-assessment objects: X axis: Individual courses, Y axis: Mean number of self-assessment objects and coverage in each course}
\label{fig:Self-assessment}
\end{figure}

\noindent\textbf{Learning by Examples and Illustrations:}
65 courses have at least one example and one illustration.
One quarter of the courses (i.e. 25) have more than 50 examples.
52 courses have been subjectively determined as low attractive ones.
Whereas, 60 courses are objectively of low attractiveness (based on the ratio of illustrations available for content units).
10 courses have been categorized as highly attractive by both criteria.

\noindent\textbf{Community Involvement:}
The content of 61 courses has been created by a single author.
Only 16 courses are the result of collaborative work.
The number of contributors are as follows:
6 courses with 2 contributors, 
3 courses with 3 contributors, 
3 courses with 4 contributors, 
2 courses with 5 contributors, 
1 course with 6 and 
1 course with 7 contributors. 
\ednote{edit}
For the rest of the courses information about their creation was not available.
The number of course reviewers has been made available for 7 courses.
Account creation was needed for more than half of the courses to comment on the course. 
The number of course downloads as well as the number of comments could not be found in most of the cases.

\noindent\textbf{Discoverability:}
For more than 60 courses, the course rank has been dramatically improved using the term ``course'' in the search keyword.
Our experiment indicates that discoverability of OCW remains low for users apart from considering good quality and clear licenses.
The number of courses w.r.t. their rank retrieved from our searched are:
14 courses with rank 1, 
7 courses with rank 2,
4 courses with rank 3,
4 courses with rank 3,
7 courses with rank between 3 and 100,
68 courses with rank above 100.


\noindent\section{Related Work}
\label{sec:background-work}

A thorough search of the literature indicates that work related to OCW quality assessment is still rather scarce.
Most of the previous works consider repositories and their impact on education rather than quality of courses.
In~\parencite{Vladoiu.2014a}, a set of quality assurance criteria is introduced considering four aspects of OCW: 1. content, 2. instructional design, 3. technology and 4. courseware evaluation.
About half of the dimensions that we consider in this work (such as availability, multilinguality) are also introduced in Vl{\u a}doiu's work.
However, some of them are not considered in this work because either they are subjective (e.g., self-containedness) or difficult to measure (e.g., relevance of the content for self-learning) or out of the scope of assessing course quality (e.g., interoperability of the interface).
Another class of related works assesses the usefulness and impact of OCW from a users's perspective, which also highly depends on quality of courses~\parencite{Chang.2014,Olufunke.2014}.
In~\parencite{Moise.2014}, a machine learning approach has been devised to support automatic OCW quality assessments.
A problem here, however, is the availability of suitable training data, which could be provided by expert sample assessments obtained using the methodology presented in this paper.

\noindent\section{Conclusion}
\label{sec:Con-Recom}

In this article, we presented a comprehensive list of quality criteria for OpenCourseWare.
We assessed a sample of 100 courses according to these criteria.
We observed that:
\begin{itemize}
\item only 28 of the courses are indeed open,
\item only 12 are available in a language other than English,
\item only 16 are available in a format facilitating reuse and re-purposeability,
\item only one third of the OCWs was updated in the last three years, and
\item less than half of the courses comprise self-assessment questions.
\end{itemize}
From our perspective, this is not a satisfactory situation.
We think the quality of OCW has to improve significantly in order to live up to its promises.
A possible solution for improving the quality is leveraging the collaboration and effort sharing on the Web.
Platforms such as SlideWiki\footnote{\url{http://slidewiki.org}} or Wikiversity\footnote{\url{http://www.wikiversity.org}}, for example, support the collaborative creation and translation of OCWs by communities of authors.

Regarding future work, we plan to further mature the quality metrics and to automate the quality assessment whenever possible.
Also collaborating with OCW repository maintainers on establishing and representing quality indicators for users appears to be an interesting development avenue.
In this regard, integrating quality assessment results into existing learning object representation standards using a standardized ontology is a further interesting aspect.

\printbibliography
\iflak
\balancecolumns

\fi

\end{document}